\documentclass[conference]{IEEEtran}
 \IEEEoverridecommandlockouts
\usepackage{cite}
\usepackage{amsmath,amssymb,amsfonts}
\usepackage{graphicx}
\usepackage{textcomp}
\usepackage{xcolor}
\usepackage{comment}
\usepackage{optidef}
\usepackage[T1]{fontenc}
\usepackage{mathtools} 
\usepackage{cuted}
\usepackage{flushend}
\usepackage{bbm}
\usepackage{dsfont}
\usepackage{mathtools}

\usepackage{algorithm}  %algorithm
\usepackage{algpseudocode}  %algorithm

\begin{document}

\title{Positioning of Multiple Unmanned Aerial Vehicle Base Stations in future Wireless Network
\thanks{The authors declare that this work has not been published in any other conference or has not been submitted for any other publication elsewhere.}}

 \author{\IEEEauthorblockN{Thushan Sivalingam\IEEEauthorrefmark{1}, K. B. Shashika Manosha\IEEEauthorrefmark{1}, Nandana Rajatheva\IEEEauthorrefmark{1}, M. Latva-aho\IEEEauthorrefmark{1}, and Maheshi B. Dissanayake\IEEEauthorrefmark{2}}
 \IEEEauthorblockA{\IEEEauthorrefmark{1}Centre for Wireless Communications, Univeristy of Oulu, Oulu, Finland }
 \IEEEauthorblockA{\IEEEauthorrefmark{2}Dept. Electrical and Electronic Engineering, Faculty of Engineering, University of Peradeniya, Sri Lanka}
 \IEEEauthorrefmark{1}\{firstname.lastname\}@oulu.fi, \IEEEauthorrefmark{2}maheshid@ee.pdn.ac.lk
 }
\maketitle

\begin{abstract}
Unmanned aerial vehicle (UAV) base stations (BSs) are reliable and efficient alternative to full fill the coverage and capacity requirements when the backbone network fails to provide such requirements due to disasters. In this paper, we consider optimal UAV-deployment problem in 3D space for a mmWave network. 
The objective is to deploy multiple aerial BSs simultaneously to completely serve the ground users. We develop a novel algorithm to find the feasible positions for a set of UAV-BSs from a predefined set of locations, subject to a signal-to-interference-plus-noise ratio (SINR) constraint of every associated user, UAV-BS's limited hovering altitude constraint and restricted operating zone constraint. We cast this 3D positioning problem as an ~$\ell_0$ minimization problem. This is a combinatorial, NP-hard problem. We approximate the ~$\ell_0$ minimization problem as non-combinatorial ~$\ell_1$-norm problem. Therefore, we provide a suboptimal algorithm to find a set of feasible locations for the UAV-BSs to operate. The analysis shows that the proposed algorithm achieves a set of the location to deploy multiple UVA-BSs simultaneously while satisfying the constraints.
%results show that our proposed algorithm an efficient location to operate while satisfying the constraints

\end{abstract}
\vspace{3mm}

\section{Introduction}
%Introduction, applications, past use cases 
%Recently, unmanned aerial vehicle (UAV) has been grabbed the attention in many fields of studies
Recently, unmanned aerial vehicles (UAVs) are expected to facilitate in many fields of studies because of their flexible attributes such as adaptive altitude, flexibility in design and movement, and mobility~\cite{8660516}. UAVs can be used as a relay to enhance coverage, capacity, and energy efficiency in wireless communication. UAVs as base stations is another application of drones which are so far used in the military for reconnaissance purposes.  
%Challenges and regulations 
%However, there are some key challenges to the usage of the UAVs such as 3D deployment in the airspace, air to ground channel modeling, network planning, flying time optimization, handover, battery life of the UAV, and performance analysis~\cite{8660516}. In addition, we need to take into account the regulations as well. Those concerns are, privacy, security, public safety, collision, and data protection. %As stated in~\cite{8660516}, according to the flying altitude UAVs can be classified into two categories, high altitude platforms (HAPs) and low altitude platforms (LAPs). HAPs are the UAVs which fly over 17 km and LAPs normally fly around few meters, maximum up to 1 km. 
%Different countries have different aviation regulations such as allowable flying altitudes as minimum distance to people, maximum distance to the airports and some areas that UAVs cannot operate due to regulation restrictions.  

%Positioning previous study

3D placement of UAV-BSs in a heterogeneous network is known to be a challenging problem~\cite{8746630,7994915,7881122,8580746}. The authors in~\cite{8746630} proposed a convex optimization based algorithm to find a position for a full-duplex UAV relay in a vehicular network. UAV based dynamic coverage in heterogeneous networks is investigated in~\cite{7994915}, where authors have proposed an optimization algorithm for UAV based floating relay cell deployment inside the existing macrocell in order to achieve dynamic and adaptive coverage. In~\cite{7881122}, a heuristic based algorithm is proposed for the 3D placement of the UAV-BSs in various geographical areas with different user densities. In~\cite{8580746}, the authors proposed drone positioning for user coverage maximization, where two techniques are proposed. First approach is the successive deployment of aerial BSs, and the second approach is the simultaneous deployment of multiple aerial BSs with $k$-mean clustering. In~\cite{8123922} an environment-aware investigation is proposed. The authors investigated ray-tracing simulations incorporated with ITU channel model and presented different channel models which mostly affect the placement of UAV-BS in real-world scenarios.

%Another study on UAV trajectory planning with jamming is investigated in~\cite{8555700}. They have proposed a convex optimization based solution for different speeds of the UAV. Besides, they have discussed static deployment scenarios as well.

The study of multiple drone cell deployment and optimization in~\cite{8287814} proposed UAV based radio access network with relays. There, user coverage, interference, and drone-to-base-station (D2B) backhaul connection features are analyzed. The optimization problem considers maximizing the user coverage while satisfying the D2B quality. Particle swarm optimization algorithm is used for approaching the solution. In~\cite{8610014}, an evolution based approach is presented for joint positioning of UAV-BS and user association. The authors proposed the method for maximizing user satisfaction by providing required data rates, and user association is conducted by using genetic algorithm and particle swarm optimization. In~\cite{8485481}, a multiple drone based positioning is proposed using exact game theory and nash equilibrium in order to perform coverage maximization and power control. In~\cite{8730845}, UAV positioning is investigated in three different approaches. First, the positions of UAVs are optimized such that the number of ground users covered by UAVs is maximized. Second, the minimum number of UAVs needed to provide full coverage for all ground users is determined. Finally, given the load requirements of the ground users, the total flight time that the UAVs need to completely serve the ground users is minimized. 

Numerous studies that have been carried out in the area of UAV positioning clearly indicate the importance of UAV positioning. Successive deployment based on linear approximation and circle packing theory are investigated in the literature. However, those algorithms have not achieved full coverage to the topology. Notwithstanding the methodologies used in the literature, it was a very important question about the aerial wireless networks, which has not been addressed yet in the literature, is to deploy multiple UAV-BSs simultaneously in order to achieve 100\% coverage to a set of ground users with target quality-of-service (QoS), while addressing the regulatory constraints. 

In this paper, we consider the problem of simultaneous deployment of multiple UAV-BSs in mmWave network. Initially, we define a set of locations for a UAV-BS to be deployed inside a rectangular geographical sub region. Next, we determine the possible combinations of predefined locations for multiple UAV-BSs in the topology. We take into consideration the mmwave transmission and multi-antenna techniques to generate directional beams to serve the users in the network. Then, we formulate a ~$\ell_0$-norm minimization problem to obtain the feasible position of the UAV-BS such that it can satisfy the QoS requirements for the ground users, limited hovering altitude constraint and restricted operating zone constraint~\cite{sivalingam2019positioning}\footnote{A pre conference version of this paper is uploaded to ~\cite{sivalingam2019positioning}.} ~\cite{jultikao19:online}\footnote{This paper is based on the research findings of the first author's master's thesis~\cite{jultikao19:online}.}. However, ~$\ell_0$-norm minimization problem is combinatorial, NP-hard. Hence, we approximate the ~$\ell_0$-norm problem as non-combinatorial ~$\ell_1$-norm problem to develop a suboptimal algorithm to find a suboptimal solution. The proposed method finds feasible positions for simultaneous deployment efficiently.

The rest of the paper is organized as follows. Section II provides the system model and the detailed analysis of the proposed algorithm. The simulations are given in Section VI, followed by conclusions in section V.

\subsection{Notation}
Boldface lowercase and uppercase letters denote vectors and matrices, respectively, and calligraphy letters denote sets. The superscript \begin{math} ^{\mbox{\scriptsize H}} \end{math} denote conjugate transpose. Complex Gaussian distribution with zero mean, variance ${\sigma^2}$ is denoted by \begin{math} \mathcal{CN} (0,{\sigma^2}) \end{math}. Finally, the absolute value of the complex number x is denoted by $|x|$ and Euclidean norm of the vector x is denoted by ${\parallel x \parallel}$.

%Equations check 
%arranging 
%Error correcting
\section{System Model and Problem Formulation}
\subsection{System Model}
Consider a multi-UAV aided mmWave wireless communication system, which consists of $D$ UAV-BSs, and each equipped with $N$ number of antennas. All UAV-BSs are represented by the set $\mathcal{D}$ = \big\{${1,2,\dots ,D}$\big\}. The location of  $j$th UAV-BS is given by $(x_j,y_j,z_j)$. $\text{UAV-BS}$ can move any direction in $xy, yz, zx$ planes to provide services to users. We assume that there is no impact on UAV-BS downlink transmission due to orientation drifts. We denote the set of all single antenna users associated with $j$th UAV-BS by $\mathcal U_j$ = \big\{${1,2,\dots ,I_j}$\big\}. We assume that all the users are randomly distributed inside a square shaped geographical region, which is bounded by the coordinates $(x_{min}^j,y_{min}^j)$ and $(x_{max}^j,y_{max}^j)$ on $xy$-plane and associated with a single UAV-BS. The location of a user \begin{math} k \in \mathcal{U}_j \end{math} is given by $(x_k^j,y_k^j,z_k^j)$. In this study, we consider only the MISO downlink wireless communication.

\subsection{Air to Ground Channel Model}
The air to ground channel model defined in~\cite{8667634}, between $j$th UAV-BS and $k$th user of the UAV-BS $j$ is given by
\begin{equation}
\textbf{h}_{j,k}^j = \sqrt{N}\sum_{p=1}^{N_P} \frac{\alpha_{k,p} \ \text{a}(\theta_{k,p})}{\sqrt{1+(d_k^j)^\gamma}},
\end{equation}
where $N_p$ is the number of multi-paths, $\alpha_{k,p}$ is the gain of $p$th path, $\theta_{k,p}$ is angle-of-departure (AoD) of the $p$th path, $\gamma$ is the path loss exponent, and $d_k^j$ is the distance between the serving UAV-BS \begin{math} j \in \mathcal{D}\end{math} and the user \begin{math} k \in \mathcal{U}_j \end{math} is given by
\begin{equation}
d_k^j=\sqrt{(x_k^j-x_j)^2+(y_k^j-y_j)^2+(z_k^j-z_j)^2}.
\end{equation}
The directive vector $\text{a}(\theta_{k,p})$ associated with AoD $\theta_{k,p}$ is defined as 
\begin{equation}
\text{a}(\theta_{k,p})= \frac{1}{\sqrt{N}}\Big[1 \ e^{-j2\pi\frac{M}{\zeta}\sin(\theta_{k,p})}\dots\ e^{-j2\pi\frac{M}{\zeta}\sin(\theta_{k,p})(N-1)}\Big]^T,
\end{equation}
where $M$ is the antenna spacing, and $\zeta$ is the wavelength of the carrier frequency. As presented in ~\cite{6834753,6363891}, mmWave channels have the non-line-of-sight (NLoS) links which are normally 20 dB weaker than the LoS link. In this study, we assume that all the users have LoS link, because UAV-BS is hovering at relatively high altitudes and having scatters around UAV-BS is very small in air interface~\cite{6834753}. Therefore, as discussed in ~\cite{7279196,7913625}, we consider only the LoS path for further study in mmWave channels. Hence, equation (1) further simplifies to~\cite{8667634}
\begin{equation}
\textbf{h}_{j,k}^j = \sqrt{N}\frac{\alpha_{k} \ \text{a}(\theta_k)}{\sqrt{1+d_{j,k}^\gamma}},
\end{equation}
where $\alpha_{k}$ is the complex gain and $\theta_k$ is the AoD of the LoS path.
\subsection{Problem Formulation}
The received signal of the $k$th user associated with $j$th UAV-BS $\textbf{r}_{j,k}$ is defined as 
\begin{multline}
\label{eqFirstOrderoffuntionG}
\textbf{r}_{j,k}=\sqrt{p_{j,k}}{(\textbf{h}_{j,k}^j)^{\mbox{\scriptsize H}}{\textbf{w}_{j,k}}{s_{j,k}}+ \sum_{\substack{i \in \mathcal{U}_j\\i\neq k}}}\sqrt{p_{j,i}}({\textbf{h}_{j,k}^j)^{\mbox{\scriptsize H}}{\textbf{w}_{j,i}}{s_{j,i}}}+ \\
\sum_{\substack{l\in \mathcal{D} \\ l\neq j}}{\sum_{\substack{{i \in \mathcal{U}_l}}}}\sqrt{p_{l,i}}{(\textbf{h}_{j,k}^l)^{\mbox{\scriptsize H}}{\textbf{w}_{l,i}}{s_{l,i}}}+n_k ,
%\vspace{-1mm}
\end{multline}
where $p_{j,k}$ is the transmit powers from $j$th UAV-BS to $k$th user, ${\textbf{h}_{j,k}^l}$ is channel vector between $k$th user of $j$th UAV-BS and $l$th UAV-BS, ${\textbf{w}_{j,k}}$ is spatial directivity of the signal sent to user $k$ from $j$th UAV-BS, and ${s_{j,k}}$ is scalar data symbol sent from $j$th UAV-BS to $k$th user. In equation (5), the first part is the desired signal that the user $k$ receives from UAV-BS $j$. The second part is interference from the UAV-BS $j$ to user $k$ while transmitting to other users. The third part is multi-user interference from other UAV-BSs to user \begin{math} k \in \mathcal{U}_j \end{math}. The fourth part is additive white Gaussian noise \begin{math} {n_k} \sim \mathcal{CN} (0,I{N_0}) \end{math}. The downlink signal-to-interference-plus-noise ratio (SINR) of the $k$th user is given by
\begin{multline}
\Gamma_{j,k} =\\
\frac{p_{j,k}|(\textbf{h}_{j,k}^j)^{\mbox{\scriptsize H}}{\textbf{w}_{j,k}}|^2}{{N_0}+\sum \limits_{\substack{i \in \mathcal{U}_j\\i\neq k}}{p_{j,i}|(\textbf{h}_{j,k}^j)^{\mbox{\scriptsize H}}{\textbf{w}_{j,i}}|^2}+\sum \limits_{\substack{l\in \mathcal{D} \\ l\neq j}}{\sum \limits_{\substack{{i \in \mathcal{U}_l}}}}{P_{l,i}|(\textbf{h}_{j,k}^l)^{\mbox{\scriptsize H}}{\textbf{w}_{l,i}}|^2}}.
\end{multline}
Figure 1 illustrates the geographical area. There are four predefined square regions, which has length R, and known number of users.

Our goal is to position the UAV-BSs one for each square in an optimum location. Let \begin{math} l_i \in \mathbb{R}^3 \end{math} be $i$th possible location a UAV-BS can operate in a predefined square region. Elements of $l_i$ are x, y, and z coordinate points, respectively. Hence the location matrix can be expressed as
\begin{equation}
L_D= \begin{bmatrix} 
x_{d1}& y_{d1} & z_{d1}\\
x_{d2}& y_{d2} & z_{d2}\\
\vdots & \vdots & \vdots\\
x_{dl} & y_{dl} & z_{dl}
\end{bmatrix}.
\end{equation}
Similarly, every UAV-BS in this network has a predefined set of locations in their respective square region. Therefore, there are many combinations of predefined locations for all UAV-BS. The total number of combinations can be expressed as 
\begin{equation}
c = {N_{pd}}^{D},
\end{equation}
where ${N_{pd}}$ is number of predefined locations inside each square, and ${D}$ represent the number of UAV-BS, in the network. %The possible combination matrix can be expressed as 
%\begin{equation}
%P_D= \begin{bmatrix} 
%L_{11}& L_{21} &\hdots& L_{D1}\\
%L_{11}& L_{21} &\hdots & L_{D2}\\
%\vdots & \vdots & \vdots &\vdots\\
%L_{1l} & L_{2l} & \hdots & L_{Dl}
%\end{bmatrix}.
%\end{equation}

According to the regulatory guidelines~\cite{8660516}, there are several concerns regarding privacy, public safety, and security. Therefore there are limiting factors for the deployment of $\text{UAV-BS}$ such as type, weight, speed, and trajectory. In this regard, we consider trajectory related constraints. Figure 1 shows the feasible and restricted regions. UAV-BS are not allowed to operate in the restricted area but need to satisfy the coverage requirements of the users inside the area. We define elliptical restricted region. Therefore, for all \begin{math} {l_i} \in \mathbb{R}^3 \end{math}, we have:
\begin{equation}
{(x-x_{di})^2}/b^2+{(y-y_{di})^2}/a^2 \geq 1
\end{equation}
where $(x_{di},y_{di})$ are the horizontal coordinates of the UAV-BS, $a$ is the major axis of the restricted area and $b$ is the minor axis of the restricted area. 
\begin{figure}[ht]
\center
\includegraphics[width=0.4\textwidth]{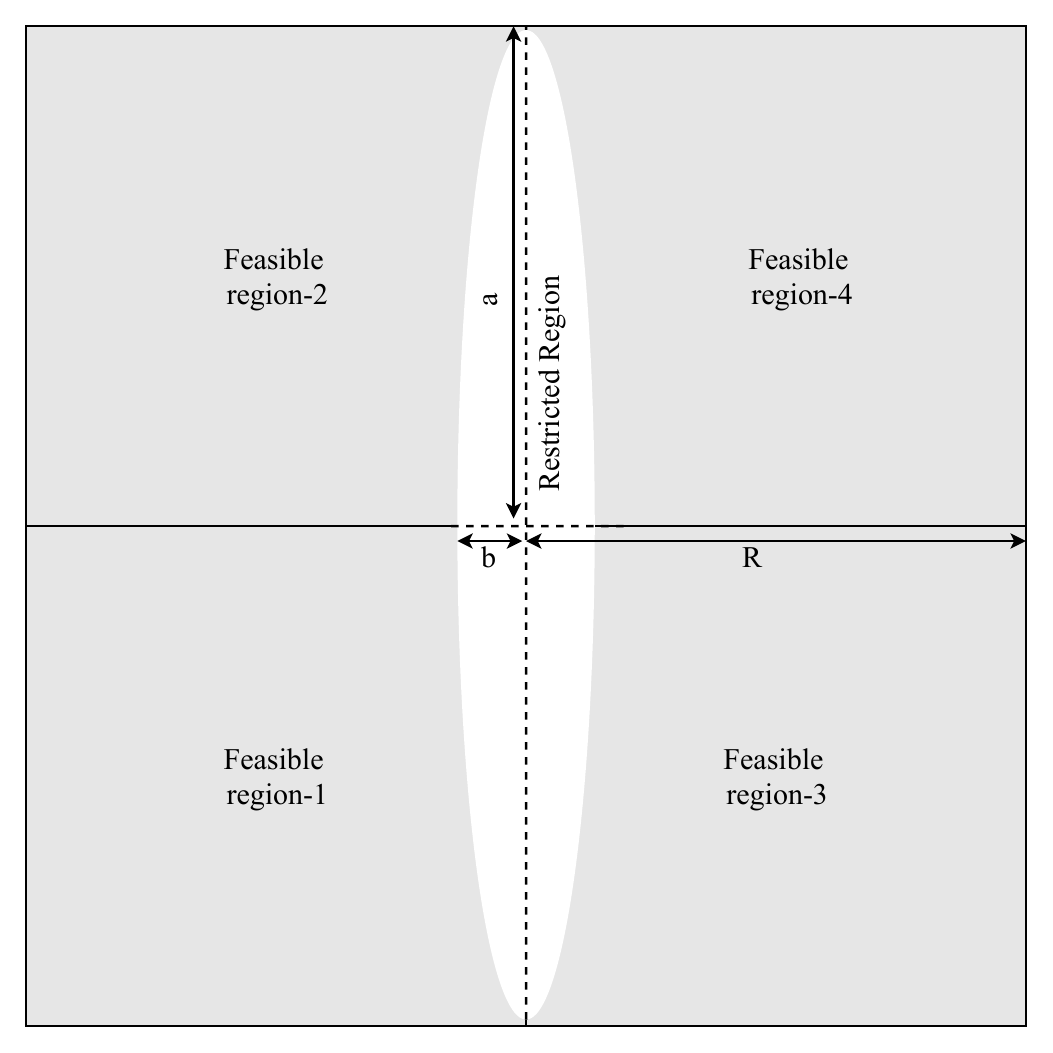}
\caption{Feasible \& restricted zones in the geographical region.}
\end{figure}

Second, we consider UAV-BS hovering-height related constraint. For the safety of the UAV-BS and regulations, each UAV-BS is assumed to have a minimum flying height of $h_{min}$. Further, UAV-BS has a maximum allowable flying height of $h_{max}$. Hence, for all \begin{math}{j} \in \mathcal{D}\end{math}, we have:
\begin{equation}
h_{min} \leq z_{di} \leq h_{max}.
\end{equation}
Let, the matrix \textbf{S} be the SINR values of all the users in all possible combinations, and \textbf{L} be the all possible combinations of UAV-BS locations. We define a vector \textbf{e}, which contains one entry equal to $1$ and all other as null. The index of the non zero entry gives the optimum combination of locations for all the UAV-BS \begin{math} j \in \mathcal{D}\end{math}. The vector \textbf{e} can expressed as, 
\begin{equation}
\textbf{e}^T= \begin{bmatrix} 
0& 0 & 0 & \dots &1 &\dots &0
\end{bmatrix}.
\end{equation}
With consideration of $\textbf{e}$, now rewrite the SINR matrix as,
\begin{equation}
\Gamma = \textbf{e}^T\textbf{S}
\end{equation}
where $\Gamma$ is a vector of all user SINR values for a possible combination. Similarly, we rewrite the possible location matrix with consideration of vector $\textbf{e}$ as,
\begin{equation}
\tau = \textbf{e}^T\textbf{L},
\end{equation}
where $\tau$ is vector of all UAV-BS positions for a possible combination. Recall that our goal is to find feasible locations for the UAVs to operate, while serving all the associated users with a guaranteed SINR and satisfying all the regulatory constraints of UAV-BS. We can formulate this optimization problem as follows: 
% \begin{mini!} 
%         {}{0}
% 		{}{}
% 		\addConstraint \Gamma_{j,k}\geq \Gamma_{th},\forall k\in \mathcal{U}_j 
% 		\addConstraint {(x-x_{di})^2}/b^2+{(y-y_{di})^2}/a^2 \geq 1, \forall {j} \in \mathcal{D}
% 		\addConstraint h_{min} \leq z_{di} \leq h_{max},\forall {j} \in \mathcal{D}
% 		\addConstraint \parallel\textbf{e}\parallel_0=1
% 		\addConstraint \textbf{e}_a  \in \big\{0,1\big\}, \ a=1,\dots, c
%   \end{mini!} 
\begin{subequations}
	\begin{align}
	\mbox{minimize } &\quad \textstyle  {0} \nonumber\\
	\mbox{subject to} &\quad  \Gamma_{j,k}\geq \Gamma_{th},\forall k\in \mathcal{U}_j \label{Optproblem_Con1.1}\\
	&\quad {(x-x_{di})^2}/b^2+{(y-y_{di})^2}/a^2 \geq 1, \forall {j} \in \mathcal{D}  \label{Optproblem_Con1.2}\\
	&\quad  h_{min} \leq z_{di} \leq h_{max},\forall {j} \in \mathcal{D} \label{Optproblem_Con1.3}  \\	
	&\quad \parallel\textbf{e}\parallel_0=1 \label{Optproblem_Con1.4}\\
	&\quad \textbf{e}_a  \in \big\{0,1\big\}, \ a=1,\dots, c	\label{Optproblem_Con1.5}	
	\end{align}
\end{subequations}
where the variable is $\textbf{e}$. The problem (14) is combinatorial and NP-hard. Hence, it requires exponential complexity to obtain the global optimum solution. Hence, we have to rely on sub-optimal algorithm to find the approximate solution to the problem. 

\subsection{Solution Approach}
We approximate the ~$\ell_0$ minimization problem (14) as a non-combinatorial ~$\ell_1$-norm problem. Therefore, we approximate constraint functions (14d) and (14e) with ~$\ell_0$ functions. The approximate problem as follows:
% \begin{mini!} 
%         {}{0}
% 		{}{}
% 		\addConstraint \Gamma_{j,k}\geq \Gamma_{th},\forall k\in \mathcal{U}_j 
% 		\addConstraint {(x-x_{di})^2}/b^2+{(y-y_{di})^2}/a^2 \geq 1,\forall {j} \in \mathcal{D}
% 		\addConstraint h_{min} \leq z_{di} \leq h_{max},\forall {j} \in \mathcal{D}
% 		\addConstraint \parallel\textbf{e}\parallel_1 \leq 1
% 		\addConstraint 0\leq \textbf{e}_a \leq 1, \ a=1,\dots, c 
% 	\end{mini!}
\begin{subequations}
	\begin{align}
	\mbox{minimize } &\quad \textstyle  {0} \nonumber\\
	\mbox{subject to} &\quad  \Gamma_{j,k}\geq \Gamma_{th},\forall k\in \mathcal{U}_j \label{Optproblem_Con1.6}\\
	&\quad {(x-x_{di})^2}/b^2+{(y-y_{di})^2}/a^2 \geq 1, \forall {j} \in \mathcal{D}  \label{Optproblem_Con1.7}\\
	&\quad  h_{min} \leq z_{di} \leq h_{max},\forall {j} \in \mathcal{D} \label{Optproblem_Con1.8}  \\	
	&\quad \parallel\textbf{e}\parallel_1 \leq 1 \label{Optproblem_Con1.9}\\
	&\quad 0\leq \textbf{e}_a \leq 1, \ a=1,\dots, c	\label{Optproblem_Con1.10}	
	\end{align}
\end{subequations}
where the variable is $\textbf{e}$. Note that binary constraints in the problem (14d) is relaxed by (15d). However, the restricted region related constraint (15b) in this problem is non-convex which makes the problem (15) a difficult one to solve. Thus we further approximate constraint (15b) in problem (15),i.e., we approximate the elliptical restriction zone by a rectangular restriction zone so that constraint (15b) become a convex one. 
\begin{figure}[ht]
\center
\includegraphics[width=0.4\textwidth]{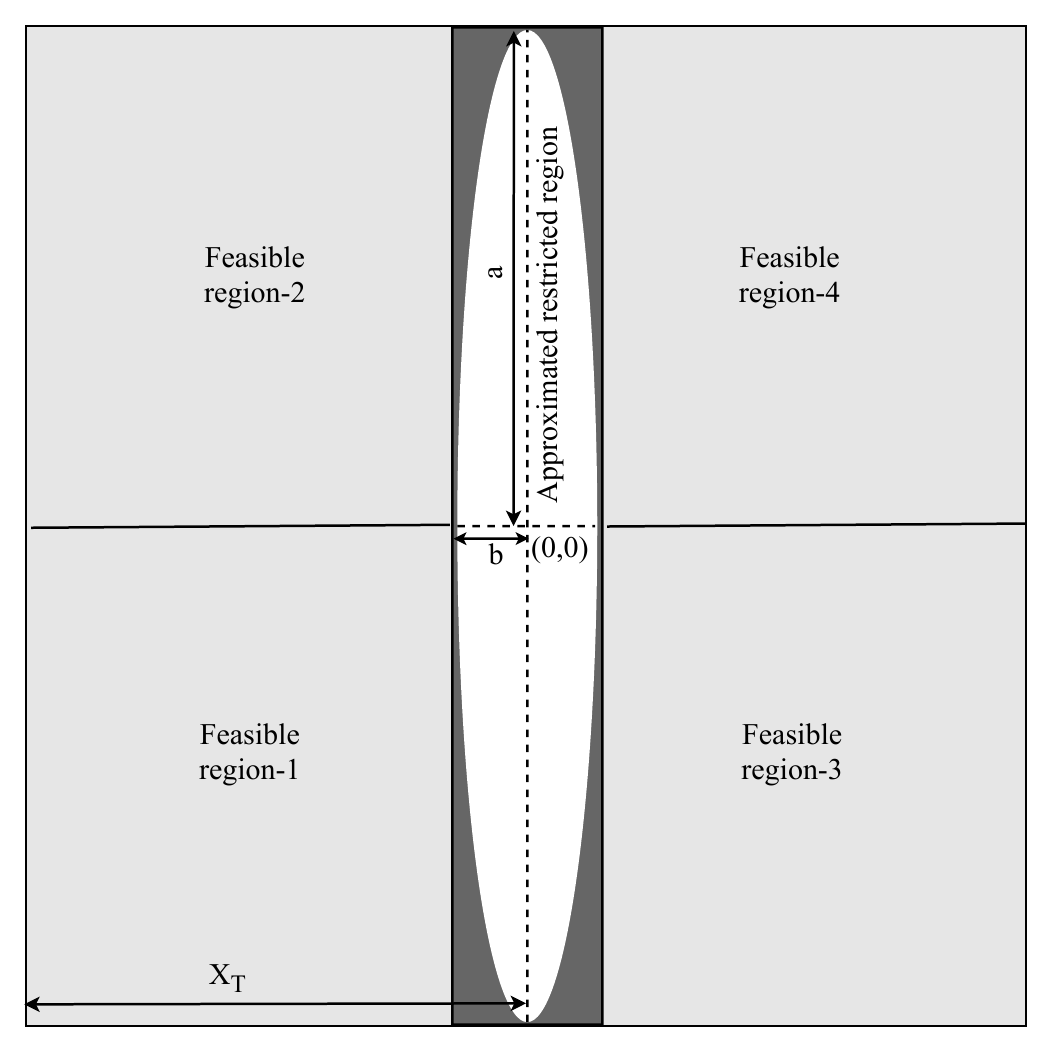}
\caption{Approximated feasible \& restricted zones in the geographical region.}
\end{figure}
We noticed that the non-convex elliptical region in the network can be approximated into linear constraints which are convex. Hence, the approximated problem can be expressed as follows:
% \begin{mini!} 
%         {}{0}
% 		{}{}
% 		\addConstraint \Gamma_{j,k}\geq \Gamma_{th},\forall k\in \mathcal{U}_j 
% 		\addConstraint x_{di}\geq X_T-b,\forall {j} \in \mathcal{D}
% 		\addConstraint x_{di}\leq -X_T+b,\forall {j} \in \mathcal{D}
% 		\addConstraint h_{min} \leq z_{di} \leq h_{max},\forall {j} \in \mathcal{D}
% 		\addConstraint \parallel\textbf{e}\parallel_1 \leq 1
% 		\addConstraint 0\leq \textbf{e}_a \leq 1, \ a=1,\dots, c 
% 	\end{mini!} 
\begin{subequations}
	\begin{align}
	\mbox{minimize } &\quad \textstyle  {0} \nonumber\\
	\mbox{subject to} &\quad  \Gamma_{j,k}\geq \Gamma_{th},\forall k\in \mathcal{U}_j \label{Optproblem_Con1.11}\\
	&\quad x_{di}\geq X_T-b,\forall {j} \in \mathcal{D}  \label{Optproblem_Con1.12}\\
	&\quad x_{di}\leq -X_T+b,\forall {j} \in \mathcal{D}  \label{Optproblem_Con1}\\
	&\quad  h_{min} \leq z_{di} \leq h_{max},\forall {j} \in \mathcal{D} \label{Optproblem_Con1.13}  \\	
	&\quad \parallel\textbf{e}\parallel_1 \leq 1 \label{Optproblem_Con1.14}\\
	&\quad 0\leq \textbf{e}_a \leq 1, \ a=1,\dots, c	\label{Optproblem_Con1.15}	
	\end{align}
\end{subequations}
where the optimization variable is $\textbf{e}$. Therefore, this convex optimization problem can be solved by standard CVX solver. The proposed algorithm to solve problem (16), i.e., to find a feasible position for a multi-UAV aided network is summarized in Algorithm 1. 
\begin{algorithm}[ht]
\textbf{1}: For a given geographical region: Define sub regions\\
\textbf{2}: Set SINR threshold, $h_{max}$, and $h_{min}$ \\
\textbf{3}: Set predefined locations inside each sub region\\
\textbf{4}: Create combination matrix for all predefined locations in the topology \\
\textbf{5}: Approximate the problem (14) by (16)\\
\textbf{6}: Find SINR values for all predefined combinations\\
\textbf{7}: Solve (16) by standard CVX solver and find $\textbf{e}$\\
\textbf{8}: Find the index of the maximum value in $\textbf{e}$, and locate the UAV-BSs\\
\caption{Multi-UAV-BS Positioning}
\end{algorithm}
	
\section{Simulation}

We consider 4 UAV-BSs to provide services to 20 single-antenna users inside a defined geographical region. The given region is equally divided into 4 square regions. All the regions contain a UAV-BS and 5 associated users. Each user can be served by only one UAV-BS. Users are assumed to be at ground level, (z = 0). We allocate equal power to all the users with beamforming. We use 20 predefined locations for the deployment in each region. Hence, there are 160,000 combinations define for the simulation. Our goal is to find the feasible set of locations out of 160,000 combinations. Table I presents the simulation parameters.

\begin{table}[ht]
\caption{Simulation Parameters}
\begin{center}
\begin{tabular}{|c|c|}
\hline
\textbf{Description}&{\textbf{Value}} \\
\hline
\hline
Size of the total area & 114 m x 114 m \\
\hline
Size of one region & 57 m x 57 m \\
\hline
Number of User in each region & 5  \\
\hline
Number of defined areas & 4  \\
\hline
Number of UAV-BSs & 4  \\
\hline
 Number of UAV-BS antenna, $N$ & 6 \\
\hline
 Path-loss exponent, $\gamma$& 2 \\
\hline
 Restricted region, b & 11 m \\
\hline
 $h_{min}$ & 22 m \\
\hline
 $h_{max}$ & 36 m \\
\hline
UAV-BS transmit power & 1 mW \\
\hline
Noise, $N_0$ & -35 dBm \\
\hline
\end{tabular}
\label{tab1}
\end{center}
\end{table}

Figure 3 shows SINR versus user index at one of the feasible location for multi-UAV-BS scenario. It can be observed that the users achieved the SINR threshold value for the ~$\ell_1$-norm minimization problem. Consider the region one and two, where the users are indexed by 1 to 5 and 6 to 10. It can be observed that the SINR value at the feasible position and SINR value at the highest altitude are the same. Therefore, the proposed algorithm chooses the highest altitude to place the first and second UAV-BSs. In the third region, both SINR values at highest (30 m) and lowest (20 m) altitudes are less than the height in between. Therefore, the algorithm chooses the position at 25 m to deploy the UAV-BS. In the fourth region, lowest altitude (20 m) has the higher SINR value than the position selected by the algorithm (25 m). This is caused by the minimum height constrain of 22 m.
\begin{figure}[ht]
    \begin{center}
\includegraphics[width=0.5\textwidth]{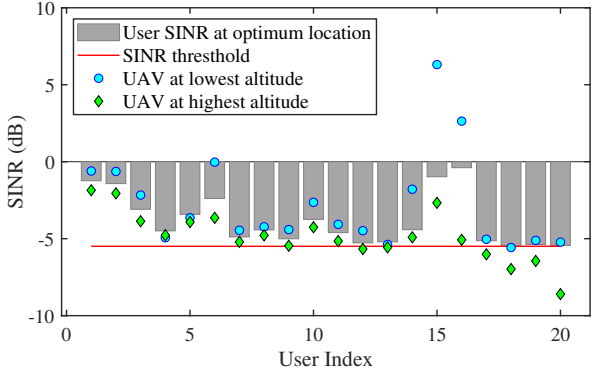}
    \end{center}
\caption{SINR versus user index at a feasible location and SINR values at highest and lowest altitudes.}
\end{figure}

Figure 4 presents the user distribution and a set of feasible positions for the UAV-BSs. Figure 4 confirms the proposed algorithm chooses feasible positions only outside the restricted region. It satisfies the operating region of constraint. 
\begin{figure}[ht]
    \begin{center}
        \includegraphics[width=0.49\textwidth]{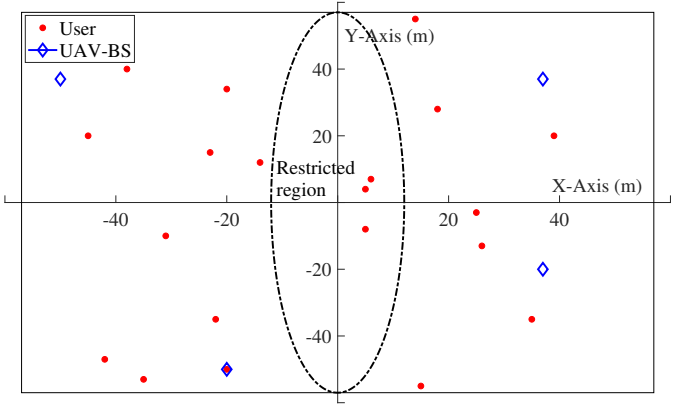}
    \end{center}
    \caption{User locations and feasible positions of the UAV-BSs, where SINR threshold = -5.49 dB, $h_{min}$ = 22 m, $h_{max}$ = 36 m, and b = 11 m.}
\end{figure}
We simulate 50 different SINR threshold values from -6.38 dB to -10 dB for the same user distribution, and plot Figure 5. We keep other constraints same during the simulation. We can observe that the feasible positions are spread outside the restricted region.
\begin{figure}[ht]
    \begin{center}
        \includegraphics[width=0.49\textwidth]{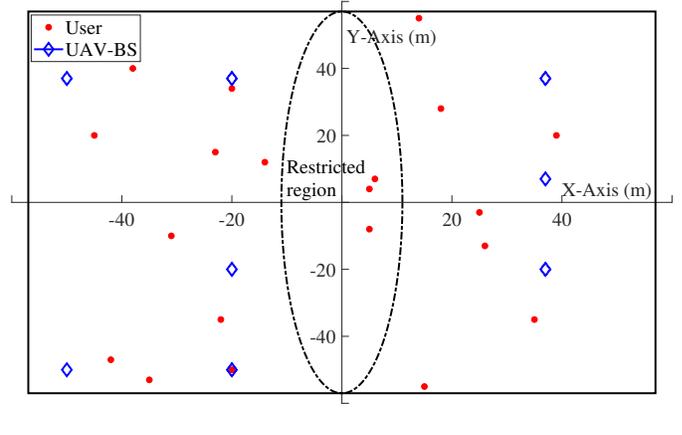}
    \end{center}
    \caption{User locations and the feasible positions of the UAV-BSs for SINR threshold values from -6.38 dB to -10 dB and same user distribution, where $h_{min}$ = 22 m, $h_{max}$ = 36 m, and b = 11 m.}
\end{figure}

We simulate 50 different user distributions for the same SINR target value, and plot Figure 6. We observed 33 user distributions achieved the SINR threshold value of -6.58 dB. Our proposed algorithm gives the same UAV-BS positions for some different user distributions. We kept the height and the restricted region constraints same during the simulation. It can be observed the feasible locations are spread outside the restricted region.
\begin{figure}[ht]
    \begin{center}
        \includegraphics[width=0.49\textwidth]{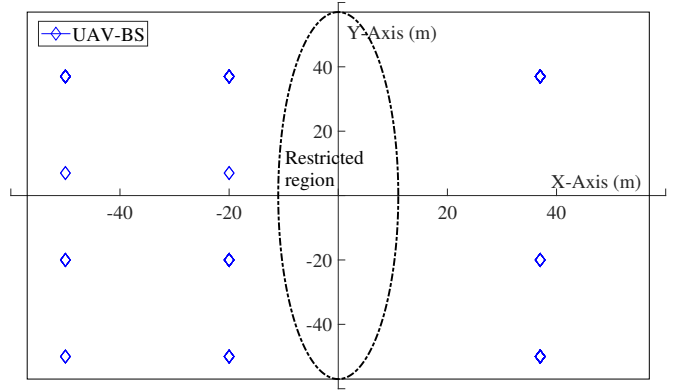}
    \end{center}
    \caption{Feasible positions of the UAV-BSs for same SINR threshold value and different user distributions, where $h_{min}$ = 22 m, $h_{max}$ = 36 m, and b = 11 m.}
\end{figure}

Figure 7 demonstrates the 3-D positions of the UAV-BSs for different user distributions. It can be observed all the possible locations are placed in between the $h_{max}$ and $h_{min}$ planes. Figure 7 confirms the proposed algorithm satisfies the height constraint. Figure 8 presents the xz-plane of the Figure 7. It can be observed that there is no UAV-BS placed inside the restricted region.
\begin{figure}[ht]
    \begin{center}
        \includegraphics[width=0.5\textwidth]{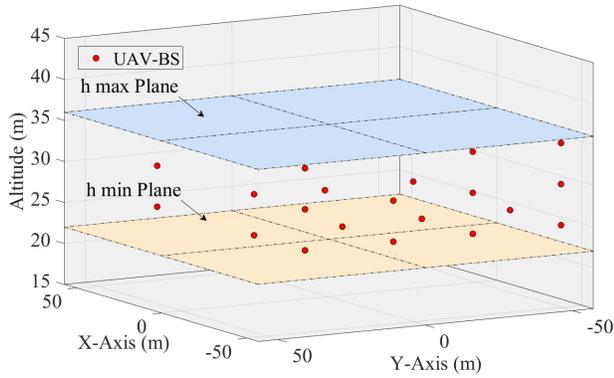}
    \end{center}
    \caption{Positions of the UAV-BSs for different user distributions in 3-D space view, where $h_{min}$ = 22 m, and $h_{max}$ = 36 m.}
\end{figure}

\begin{figure}[ht]
    \begin{center}
        \includegraphics[width=0.5\textwidth]{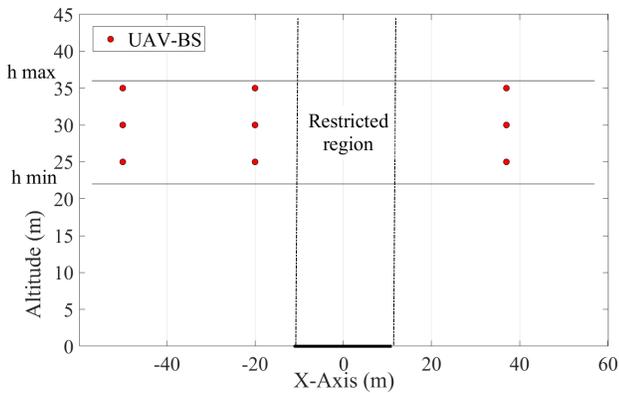}
    \end{center}
    \caption{Positions of the UAV-BSs for different user distributions in xz-plane view.}
\end{figure}

In summary, these results show that the proposed algorithm gives a set of feasible positions to deploy multiple UAV-BSs simultaneously. Figure 3 shows the SINR values of the ground users at one of the feasible locations. Meantime, the respective deployment of the UAV-BSs present in Figure 4. Besides, Figure 5 demonstrates the set of UAV-BS positions for the same user distribution and different SINR threshold values. Collectively, the above results satisfy our SINR and restricted region constrains. Furthermore, Figure 6 illustrates the set of UAV-BS positions for different user distributions and the same SINR threshold value. The 3-D space view, xz-plane view of Figure 6 presents in Figure 7 and Figure 8, respectively, and we can observe our algorithm achieves the height and restricted region constraints. Overall, these findings will doubtless be matched with all constraints mentioned in the problem (16). 

\section{Conclusions}
In this paper, we investigate how simultaneous 3D deployment of multiple UAV-BSs can be introduced in the mmWave wireless network. In particular, we focus on finding a feasible location within a sub region by addressing QoS and regulatory constraints. The objective is to deploy multiple aerial BSs simultaneously to completely serve the ground users. First, by using a set of predefined locations for each UAV-BS in every sub region, we have created the combination matrix for all predefined locations in the topology and derived the SINRs for all the possible locations for the UAV-BSs. Second, to find the feasible locations of the UAV-BSs, We have formulated a ~$\ell_0$ norm minimization problem, which is combinatorial, NP-hard. This ~$\ell_0$ norm minimization problem has been approximated as a non-combinatorial ~$\ell_1$ norm minimization problem. Then, we have proposed a suboptimal algorithm to solve it. Simulation results considering regions with the same and different user distributions confirmed the simultaneous deployment of the proposed algorithm for different SINR threshold values while addressing the regulatory constraints.

\bibliographystyle{IEEEbib}
\bibliography{main.bib}

\begin{thebibliography}{10}

\bibitem{8660516}
M.~{Mozaffari}, W.~{Saad}, M.~{Bennis}, Y.~{Nam}, and M.~{Debbah},
\newblock ``A tutorial on uavs for wireless networks: Applications, challenges,
  and open problems,''
\newblock {\em IEEE Communications Surveys Tutorials}, pp. 1--1, 2019.

\bibitem{8746630}
P.~{Pourbaba}, K.~B.~S. {Manosha}, S.~{Ali}, and N.~{Rajatheva},
\newblock ``Full-duplex uav relay positioning for vehicular communications with
  underlay v2v links,''
\newblock in {\em 2019 IEEE 89th Vehicular Technology Conference
  (VTC2019-Spring)}, April 2019, pp. 1--6.

\bibitem{7994915}
Y.~{Li} and L.~{Cai},
\newblock ``Uav-assisted dynamic coverage in a heterogeneous cellular system,''
\newblock {\em IEEE Network}, vol. 31, no. 4, pp. 56--61, July 2017.

\bibitem{7881122}
E.~{Kalantari}, H.~{Yanikomeroglu}, and A.~{Yongacoglu},
\newblock ``On the number and {3D} placement of drone base stations in wireless
  cellular networks,''
\newblock in {\em 2016 IEEE 84th Vehicular Technology Conference (VTC-Fall)},
  Sep. 2016, pp. 1--6.

\bibitem{8580746}
J.~{Sun} and C.~{Masouros},
\newblock ``Drone positioning for user coverage maximization,''
\newblock in {\em 2018 IEEE 29th Annual International Symposium on Personal,
  Indoor and Mobile Radio Communications (PIMRC)}, Sep. 2018, pp. 318--322.

\bibitem{8123922}
I.~{Bor-Yaliniz}, S.~S. {Szyszkowicz}, and H.~{Yanikomeroglu},
\newblock ``Environment-aware drone-base-station placements in modern
  metropolitans,''
\newblock {\em IEEE Wireless Communications Letters}, vol. 7, no. 3, pp.
  372--375, June 2018.

\bibitem{8287814}
W.~{Shi}, J.~{Li}, W.~{Xu}, H.~{Zhou}, N.~{Zhang}, S.~{Zhang}, and X.~{Shen},
\newblock ``Multiple drone-cell deployment analyses and optimization in drone
  assisted radio access networks,''
\newblock {\em IEEE Access}, vol. 6, pp. 12518--12529, 2018.

\bibitem{8610014}
J.~{Plachy}, Z.~{Becvar}, P.~{Mach}, R.~{Marik}, and M.~{Vondra},
\newblock ``Joint positioning of flying base stations and association of users:
  Evolutionary-based approach,''
\newblock {\em IEEE Access}, vol. 7, pp. 11454--11463, 2019.

\bibitem{8485481}
L.~{Ruan}, J.~{Wang}, J.~{Chen}, Y.~{Xu}, Y.~{Yang}, H.~{Jiang}, Y.~{Zhang},
  and Y.~{Xu},
\newblock ``Energy-efficient multi-uav coverage deployment in uav networks: A
  game-theoretic framework,''
\newblock {\em China Communications}, vol. 15, no. 10, pp. 194--209, Oct 2018.

\bibitem{8730845}
A.~{French}, M.~{Mozaffari}, A.~{Eldosouky}, and W.~{Saad},
\newblock ``Environment-aware deployment of wireless drones base stations with
  google earth simulator,''
\newblock in {\em 2019 IEEE International Conference on Pervasive Computing and
  Communications Workshops (PerCom Workshops)}, March 2019, pp. 868--873.

\bibitem{sivalingam2019positioning}
Thushan Sivalingam, K.~B.~Shashika Manosha, Nandana Rajatheva, M.~Latva-aho,
  and Maheshi~B. Dissanayake,
\newblock ``Positioning of multiple unmanned aerial vehicle base stations in
  future wireless network, 2019,'' [Online].Avilable:
  {https://arxiv.org/abs/1911.03929}.

\bibitem{jultikao19:online}
S.Thushan,
\newblock ``Positioning of multiple unmanned aerial vehicle base stations in
  future wireless network, {M}aster's thesis, {U}niversity of {O}ulu,
  {F}inland, 2019,'' [Online].Avilable:
  {http://jultika.oulu.fi/Record/nbnfioulu-201908132758}.

\bibitem{8667634}
N.~{Rupasinghe}, Y.~{Yapıcı}, İ. {Güvenç}, M.~{Ghosh}, and Y.~{Kakishima},
\newblock ``Angle feedback for noma transmission in mmwave drone networks,''
\newblock {\em IEEE Journal of Selected Topics in Signal Processing}, vol. 13,
  no. 3, pp. 628--643, June 2019.

\bibitem{6834753}
M.~R. {Akdeniz}, Y.~{Liu}, M.~K. {Samimi}, S.~{Sun}, S.~{Rangan}, T.~S.
  {Rappaport}, and E.~{Erkip},
\newblock ``Millimeter wave channel modeling and cellular capacity
  evaluation,''
\newblock {\em IEEE Journal on Selected Areas in Communications}, vol. 32, no.
  6, pp. 1164--1179, June 2014.

\bibitem{6363891}
T.~S. {Rappaport}, E.~{Ben-Dor}, J.~N. {Murdock}, and Y.~{Qiao},
\newblock ``38 {GHz} and 60 {GHz} angle-dependent propagation for cellular
  peer-to-peer wireless communications,''
\newblock in {\em 2012 IEEE International Conference on Communications (ICC)},
  June 2012, pp. 4568--4573.

\bibitem{7279196}
G.~{Lee}, Y.~{Sung}, and J.~{Seo},
\newblock ``Randomly-directional beamforming in millimeter-wave multiuser miso
  downlink,''
\newblock {\em IEEE Transactions on Wireless Communications}, vol. 15, no. 2,
  pp. 1086--1100, Feb 2016.

\bibitem{7913625}
D.~{Zhang}, Z.~{Zhou}, C.~{Xu}, Y.~{Zhang}, J.~{Rodriguez}, and T.~{Sato},
\newblock ``Capacity analysis of noma with mmwave massive mimo systems,''
\newblock {\em IEEE Journal on Selected Areas in Communications}, vol. 35, no.
  7, pp. 1606--1618, July 2017.

\end{thebibliography}

\end{document}